\documentstyle[prb,aps,floats,epsf]{revtex}

\begin{document}

\draft
\preprint{}

\twocolumn[\hsize\textwidth\columnwidth\hsize\csname@twocolumnfalse%
\endcsname

\title{Periodic magnetoconductance fluctuations in triangular
       quantum dots in the absence of selective probing}
\author{P.~B\o ggild, A.~Kristensen, H.~Bruus,
        S.M.~Reimann, and P.E.~Lindelof}
\address{Niels Bohr Institute, \O rsted Laboratory, University of
Copenhagen, Universitetsparken 5, DK-2100 Copenhagen \O, Denmark}
\date{Phys.\ Rev.\ B {\bf 57}, 15408 (1998)}
\maketitle
\begin{abstract}
We have studied the magnetoconductance of
quantum dots with triangular symmetry and areas down to
0.2~$\mu$m$^2$, made in a high mobility
two-dimensional electron gas embedded in a GaAs-Al$_{x}$Ga$_{1-x}$As
heterostructure.  Semiclassical simulations show
that the gross features in the measured magnetoconductance are caused by
ballistic effects.  Below 1~K we observe a
strong periodic oscillation, which may be explained
in terms of the Aharanov-Bohm flux quantization through the area of a
single classical periodic orbit.  From a numerical and analytical
analysis of possible trajectories in hard- and soft-walled
potentials, we identify this periodic orbit as the enscribed
triangle. Contrary to other recent experiments, this orbit is not
accessible by classical processes for the incoming collimated beam.
\end{abstract}

\pacs{}
]

\section{Introduction}
\label{sec:introduction}
Over the past two decades, the technological advances of device
fabrication and semiconductor growth techniques have made possible
the studies of small and impurity-free electron systems.
At sufficiently low temperatures both the elastic mean free path and
the phase coherence length become larger than the characteristic
length scales of the sample, and its transport
properties\cite{MesoscopReview} can reveal pronounced quantum
interference effects.  A device, particularly well suited for studies
of these effects, is the quantum dot\cite{QdotReview} created by
electrostatical lateral confinement of a
two-dimensional electron gas (2DEG). The size and shape of the dot can
be changed by varying the voltage of the electrodes, which determine
the confining potential, thus providing a possibility to modify the
system {\it in situ}.
In ballistic quantum dots, where the boundary of the confining
potential has a shape, that generates chaotic classical dynamics,
one observes random, reproducible conductance fluctuations, provided
the Fermi wavelength $\lambda_{\rm F}$ is small compared to the dot
size \cite{MesoscopReview}.  For small dots, where this requirement is
not quite fulfilled, the wave nature of the electrons reduces the
sensitivity to the initial conditions characteristic of chaotic
dynamics, and a more regular electronic motion appears.
Recently, the attention has turned towards
quantum dots that are just marginally chaotic, i.e.  devices capable
of showing both regular and chaotic behavior described by a mixed
phase space
\cite{QdotReview,Marcus92a,Chang94a,Bruus94a,Persson95a,Bird96a,Zozoulenko97a,Christensson97a}.

Besides the size and overall shape of the potential, the leads feeding
current into the quantum dot are often perturbations on an
intentionally regular confining potential. As the leads are gradually
opened, the dynamics therefore changes from regular to chaotic\cite{Bird95b}.
Furthermore, it is expected theoretically\cite{Buettiker86a}
that the total phase breaking rate $\tau_{\rm \phi}^{-1}+\tau_{\rm
esc}^{-1}$ is enhanced, as escaped electrons are re-injected with
uncorrelated phases. As the leads are opened, the quantum dot is
no longer an isolated system, its eigenstates become life-time
broadened, and in the limit of wide open leads they eventually evolve
into scattering resonances in a continuous spectrum.  In transport
measurements this allows only the gross shell structure rather
than individual levels to be resolved\cite{Persson95a}.

Several observations of periodic magnetoconductance fluctuations in
open dots have been interpreted semiclassically, based on periodic
orbits, or quantum-mechanically, based on scarred wavefunctions
\cite{QdotReview,Marcus92a,Chang94a,Persson95a,Bird96a,Zozoulenko97a,Christensson97a}.
The observed
periods $\Delta B$ in magnetic field can be related to the area $F$
encompassed by a periodic orbit using the Aharonov-Bohm type relation
$F\Delta B = \Phi_0$, where $\Phi_0 = h/e$ is the flux quantum.
This relation and generalizations thereof can be more rigorously
obtained from semiclassical periodic orbit theory\cite{SemiclReview}.
In the spirit of Bohr and Sommerfelds quantization rule, periodic
orbit theory provides a connection between the classical action of
the periodic orbits and the density of states in the quantum regime.

When analyzing the experimental results, however, the resolution
seldom warrant an interpretation in terms of the full periodic orbit
theory: to extract the quantum density of states from the transport
measurements is a difficult, if not impossible task.
This is also the case for our work, so to explain the quasi-periodic
magnetoconductance fluctuations observed for our triangular dots, we
simply start from the classical action of
a periodic electron orbit of length $L$ encompassing an area $F$,
which for a constant wavelength $\lambda_{F}$ becomes

\begin{equation} \label{BS}
N = \frac{1}{h} \oint \! ({\bf p} -e{\bf A}) \!\cdot\! d{\bf q} =
\frac{L}{\lambda_{\rm F}} - \frac{BF}{\Phi_{0}}.
\end{equation}
This yields a magnetic field dependent quasi-period $\Delta B(B)$
given by

\begin{equation} \label{DeltaB}
\Delta B(B) = \left|\frac{dN}{dB}\right|^{-1},
\end{equation}
since both $L$ and $F$ depend on the magnetic field. In most
experiments such an oscillation has to be extracted from a complicated
background consisting of additional periodic components as well as an
aperiodic part\cite{Marcus92a}.

The role of leads as selective probes of resonant states has recently
been emphasized\cite{Bird96a,Zozoulenko97a,Christensson97a}.  The
collimated beam of electrons injected through one lead selects a
set of momenta and coordinates, i.e.\ a
particular part of phase space. Semiclassically, resonances occur when
the direct trajectories are being injected close to
periodic orbits in phase space.

In this work we focus on the role of leads for providing a mechanism
to select periodic orbits in quantum dots with mixed dynamics, and we
have chosen a triangular geometry of the dot to obtain a particularly
simple set of periodic orbits. In
Sec.~\ref{sec:experiment} we describe the experiment and a simple
geometrical model for the electrostatic potential induced by the
gates. Then, in  Sec.~\ref{sec:classical} we use classical
simulations to assert that the periodic fluctuations in the
low temperature magnetoconductance are not of classical origin.
In Sec.~\ref{sec:quantum} we
compare the measured quasi-period of the quantum oscillations to the
results of a numerical and analytical analysis of the classical
trajectories. We show that particularly robust and strong conductance
oscillations can be related to a single orbit that is classically
inaccessible from the leads. Finally, in Sec.~\ref{sec:discussion}
we discuss the origin and the implications of the small coupling
between this periodic orbit and the leads. It is concluded that
selective probing is not necessary to detect periodic orbits in a
small, open cavity with mixed dynamics.

\section{The experiment}
\label{sec:experiment}

A GaAs-Al$_{x}$Ga$_{1-x}$As heterostructure with a 2DEG embedded 90 nm
under the surface is used as a starting point for the device fabrication.
The mobility of the 2DEG after processing is about 200~m$^2/$Vs and
the carrier density  $1.7\!\times\!10^{15}$~m$^{-2}$.  Using
electron-beam lithography, three hexagonal aluminium gates are
deposited on the surface of the heterostructure, as shown in the
insets of Fig.~\ref{oscillations}. By applying a negative gate voltage
of $-0.3$~V, an equilateral triangular cavity with open corners is
formed between the gates.  We present measurements performed on three
samples, in the following referred
to as A1, A2 and B1. Sample A1 and A2 are nominally identical
while sample B1 has the same lithographic shape, but an area twice as
large. The lithographic width
$W$ of the leads is 385~nm for sample B1 and 269~nm for sample A1 and
A2. The corresponding pinch-off voltages for the three samples are
$V_{\rm B1}=-1.0$~V, $V_{\rm A1}=-0.50$~V, and $V_{\rm
A2}=-0.57$~V. The sidelength of the hexagonal gates is $2W$ in all
samples. Based on the classical ballistic features described in
Sec.~\ref{sec:classical}), we find the carrier densities to range
between 0.7 and 1.0 in units of $10^{15}$~m$^{-2}$  for all confined
quantum dots, corresponding to a Fermi wavelength  $\lambda_{\rm
F}\approx 80$~nm.

The electrostatic potential changes the area and shape of the dot
significantly from the first formation to pinch-off.
In the analysis of our data,
we therefore need to estimate the size of the dot as a function of
gate voltage. When the dot is formed at the gate voltage $V_{\rm
d}=-0.3$~V, the edge of the electron gas is assumed to be directly
under the edge of the gates. As the gate voltage is made more
negative, a region of width $d$ is depleted around the gates, until
the three point contact regions are fully depleted at the
pinch-off voltage $V_{\rm p}$. In a theoretical study of a half plane
gate situated in the plane of a 2DEG, Shikin and Larkin\cite{Larkin90a}
found a linear relation between the depletion width $d$ and the gate
voltage $V_{\rm g}$. The depletion zone around one of the hexagonal
gates is assumed to be independent of the two other gates, and
consequently the dot shrinks linearly from the lithographic shape at
$V_{\rm g}=V_{\rm d}$ to a smaller triangle near pinch off at $V_{\rm
g}=V_{\rm p}$. We can now express the area $F$ of the dot as a
function of the gate voltage, or more conveniently as a function of
the dimensionless gate voltage $\tilde{V} = (V_{\rm g} -
V_{\rm d})/(V_{\rm
p} - V_{\rm d})$ as follows: $F_{\rm dot} \approx  W^2 (\sqrt{3}/4)
[(4-3\tilde{V}/2)^2 - 3(1- \tilde{V})^2]$. With this model we estimate
the areas 0.41~$\mu$m$^2$ (0.83~$\mu$m$^2$) for sample A1 (B1)
at the formation of the dots, and roughly half these areas near pinch-off.

The samples are cooled to 20~mK in a dilution fridge located in a
RF-shielded room. Using standard lock-in techniques the three-terminal
devices are measured in a current-biased two-probe configuration with
the third probe floating.  Excitation currents
are typically 1~nA at an ac-frequency of 117~Hz.  Despite filtering,
the electron gas is not cooled below 50~mK, as estimated from
the activated behavior of the resistivity $\rho_{xx}$ on the $\nu =
2/3$ fractional quantum Hall plateau and from the temperature
dependence of the low field conductance fluctuations. The magnetic
field sweep rate $dB/dt$ was kept below 0.15~T/min to avoid additional
heating due to eddy currents.

\section{Classical effects}
\label{sec:classical}

\begin{figure}[t]
\epsfysize=80mm \centerline{\epsfbox{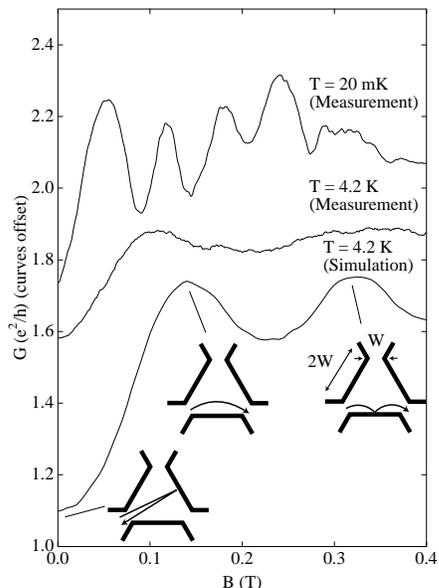}}
\caption{
\label{simulations}
Magnetoconductance of sample A1 at $T=20$~mK (top) and  $T=4.2$~K
(middle) compared to the result of a classical simulation
involving 20000 injected particles per data point (bottom), with 
energies distributed according to the temperature $T=4.2$~K.  The model 
potential is a step potential shaped like the gates but softened by a 
Gaussian function of width $\gamma = 45$~nm.  The carrier density is 
$0.7\!\times\!  10^{15}$~m$^{-2}$. We note that the same choice of
parameters yields the fit of the quasi-periods in
Fig.\ \protect\ref{fielddependence}c. The insets schematically show the
origin of the variations: the main dip is related to backscattered
trajectories, whereas the two conductance maxima correspond to direct
transmission and a single skipping orbit.}
\end{figure}

Before studying the quantum interference effects, we first have to
identify those parts of the conductance fluctuations, which are related
to purely classical effects.  One possible way to discern classical
conductance variations from quantum fluctuations is to increase the
temperature.  The classical variations are far less temperature
dependent than the quantum fluctuations, reflecting the elastic mean
free path being less temperature dependent than the phase-breaking
length\cite{MesoscopReview,QdotReview}.  Above 1~K the quantum
fluctuations are largely suppressed, and we are left with the
classical variations, which for sample A1 near pinch-off are
dominated by a large conductance dip around zero field as shown in
Fig.~\ref{simulations}.  This zero field dip, which can be quite large
(of the order of  $2e^2/h$), is accompanied by two smaller conductance
maxima at 0.11~T
and 0.35~T. We find this overall shape and field scale of the
magnetoconductance preserved up to about 10~K, where the variations
vanish due to the increased scattering\cite{Boeggild96a}.  Similar
behavior is observed for samples A2 and B1.

To identify the classical ballistic part of the conductance, we
performed a semiclassical simulation in a potential resembling the
effective potential of the cavity, and compared the
calculated conductance to low and intermediate temperature
measurements. The advantage of this approach is twofold: it allows us
to gain an understanding of the physics behind the conductance
variations in the classical regime, and it provides a way to make
consistent deductions on the shape of the effective potential and the
carrier density in the dot, as these two parameters determine the
amplitude, shape and field scale of the magnetoconductance
variations. The four steps of our analysis are straightforward. 1) A
soft model potential is defined on a lattice, allowing to calculate
the force on each point in the $xy$ plane. 2) An ensemble of electrons
is injected with cosine distributed angles in lead $i=1$, and
tracked to the exit leads $j=1,2,3$ using the classical equations of
motion.  3) The transmission coefficients $T_{ji}$ are calculated as
the number of modes in lead $i$ times the probability of
ending in lead $j$ when starting from lead $i$.  4) Using
Landauer-B\"uttiker formalism for a three probe dot with triangular
symmetry\cite{Marcus93a}, the conductance becomes $G_{12,12} =
\frac{2e^2}{h}(T_{21} + T_{31}^2/[T_{21}+T_{31}])$, where the
triangular symmetry conditions $T_{12}=T_{23}=T_{31}$ and $T_{13} =
T_{21} = T_{32}$ are used\cite{Boeggild96a}.

We have used two model potentials. Firstly, the three-fold
H\'enon-Heiles potential\cite{SemiclReview,Brack93a}
with appropriately chosen parameters,
which in polar coordinates has the form $U(r,\theta) =
\alpha r^2 + \beta r^3 \sin(3\theta)$; and secondly, a gate-shaped
step function ($U = V_{\rm g}$ on the gates and $U=0$ between the
gates), convoluted by a Gaussian function of width $\gamma$ to emulate
the soft walls characteristic of electrostatic
potentials\cite{Larkin90a}. By changing the smoothing width $\gamma$
the convoluted potential can be varied  from being nearly
hard-walled, $\gamma \simeq 0$, to being very soft-walled, $\gamma \gg
W$. For $0<\gamma<W/2$ the potential is fairly hard-walled with
corresponding flat regions in the center at distances roughly greater
than $\gamma$ away from the gates. In contrast, the H\'enon-Heiles
potential always has a soft parabolic shape in the center.

In Fig.~\ref{simulations} we compare the magnetoresistance traces of
sample A1 close to pinch-off at temperatures $T=4.2$~K and $T=0.02$~K
with a numerical 4.2~K simulation.
For each calculated value $G(B)$, 
20000 trajectories with energies distributed according to the finite
temperature, are started outside the cavity, the majority being
backscattered before entering the dot. We find fair agreement
between measurement and simulation by choosing rather hard walls
with a smoothing width $\gamma$ of only
45~nm, and by choosing a carrier density $n = 0.7\!\times\!
10^{15}$~m$^{-2}$. The measured conductance variations in
Fig.~\ref{simulations} are smaller than observed in the simulation,
which we believe is due to temperature-induced scattering processes
not taken into account; at 8~K the classical variations are nearly
washed out. The simulation could probably be further improved using a
self-consistent potential such as in
Ref.~\onlinecite{Stopa96a}. However, the same values used for $\gamma$
and $n$ in the simple simulation presented in 
Fig.~\ref{simulations} also lead to an excellent agreement for the
quasi-period as shown in Fig.~\ref{fielddependence}c to be discussed
in Sec.~\ref{sec:quantum}. We therefore conclude that the simulation
works satisfactorily, and it shows that the conductance dip around
$B=0$~T is caused by trajectories reflected by the flat wall opposing
the source lead.  As the magnetic field is increased to $B
\approx$~0.1~T, the collimated electron beam is directed into the exit
lead, leading to a maximum in the conductance.  A similar peak is seen
at $B \approx$~0.3~T, where the trajectories perform one bounce on the
edge of the potential before exiting.  Similar geometric resonances
have been studied in semiconductor double slits\cite{Houten89a},
in rectangular cavities\cite{Zozoulenko97a,Boeggild96c} and in etched
triangular cavities\cite{Christensson97a}.

We conclude that the slow variations in the magnetoconductance at
intermediate temperatures are of classical nature containing
no oscillatory structure on scales smaller than 0.1~T. We have
estimated $n$ to be between 0.7 and 1.0 in units of
$10^{15}$~m$^{-2}$ and found evidence for the effective potential
being fairly hard-walled and therefore quite far from the
H\'enon-Heiles type.  As we now have accounted for the conductance
features seen at intermediate temperatures we now proceed to the low
temperature regime.

\section{Quantum interference effects}
\label{sec:quantum}

\begin{figure}[h]
\epsfxsize=\columnwidth \centerline{\epsfbox{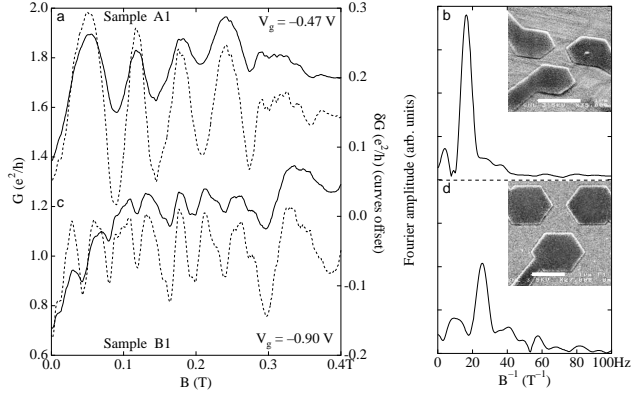}}
\caption{
\label{oscillations}
At low temperatures the magnetoconductance traces exhibit an 
oscillatory structure up to roughly 0.3~T (sample B1) and 0.4~T 
(sample A1), which is considerably stronger than random conductance 
fluctuations.  Panel (a) shows for sample A1 the raw conductance
measurement (full curve) and the measurement with the classical
background filtered out (dashed curve), while panel (b) shows the 
corresponding power spectrum ($0 < B < 0.3$ T).  
Accordingly, panel (c) and (d) show conductance and 
power spectrum for sample B1.  For the small dot A1, the oscillations 
are almost free of other harmonics, whereas sample B1 exhibits more 
frequencies.  For sample A1, the oscillations die out at higher fields 
than for sample B1, and persist over larger range of gate voltage.  
The insets in the power spectra show SEM micrographs of the samples 
with white bars of length 1~$\mu$m.}
\end{figure}

In all samples we find pronounced oscillatory structure in the 
magnetoconductance below 1~K in addition to the slow variations at 
higher temperatures.  As seen in Fig.~1 the low temperature 
conductance oscillations in sample A1 completely dominates the 
classical conductance variations, making the classical background 
difficult to observe.  In Fig.~\ref{oscillations} the 
magnetoconductance data of samples A1 and B1 are shown before (full 
lines) and after (dashed lines) applying a high pass filter to remove 
the relatively slow classical variations.

The oscillations in the smaller samples of type A are generally twice 
as strong as in the larger samples of type B, they also persist over 
larger ranges of gate voltages, and they tend to be dominated by a 
single quasi-period with little other structure.  The main oscillation 
of sample B1 is accompanied by other periodic and aperiodic 
components.  The dominating periods measured in sample A1 (40~mT~-- 
70~mT) are larger than those in sample B1 (25~mT~-- 40~mT).  The peak 
positions generally move as the size of the quantum dot is changed, 
i.e.  by varying the gate voltage.  We point out that the samples A1 
and A2 are not identical, and that the same sample is different from 
cool-down to cool-down. The overall behaviour is however conserved.
The oscillations continue with roughly constant quasi-period up to
about 0.3~T in all three samples, as reflected in the narrow peaks in
the power spectra shown in Figs.~\ref{oscillations}b
and~\ref{oscillations}d. Above this field, the fluctuations become
weaker while the quasi-period (peak-to-peak distance)
increases. In all cases the oscillations remain essentially unchanged
on a time scale of 24~hours, except for minor differences in shape due
to occasional redistribution of the potential, presumably caused by
charge hopping between the donor atoms in the GaAlAs barrier
layer\cite{Stopa96a}.

\begin{figure}[h]
\epsfysize=80mm \centerline{\epsfbox{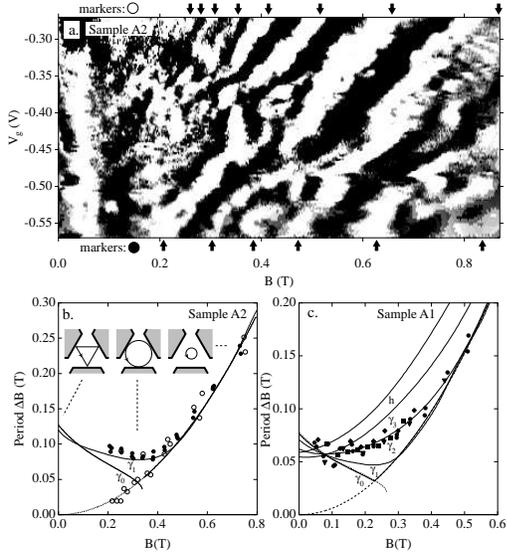}}
\caption{
\label{fielddependence}
Panel (a) shows the conductance of sample A2 in a gray-scale plot
(bright denotes low conductance), versus gate voltage and magnetic
field.  The gate voltage ranges from $-0.27$~V (formation point),
to the pinch-off voltage $-0.57$ V for this particular dot.
The arrows at the top of the graph show the conductance maxima being
periodic in $B^{-1}$ characteristic of Landau Level (LL) oscillations
in a electron system with carrier density of $1.0\!\times\!  10^{15}$
m$^{-2}$.  The arrows at the bottom show the $B$ quasi-periodic
oscillations related to the triangular loop orbit, with an increasing
quasi-period at high fields. In panel (b) the measured quasi-periods of
sample A2 are compared to the analytical expression
Eq.~(\protect\ref{period}) for a hard wall potential~($\gamma_{0}$), as
well as numerical calculations for a step potential with soft region
width of 15~nm ($\gamma_{1}$). The best fit is obtained by choosing
the carrier density $n_{\rm A2} = 1.0\!\times\!~10^{15}$ m$^{-2}$.
The white markers show the measured oscillation quasi-periods of
the sample at the point of formation being $1/B$ periodic, down to much
lower fields than when the dot is near pinch-off (black markers).  
Similarly, in panel (c) results for the quasi-period of sample A1 is
compared to the analytical expression Eq.~(\protect\ref{period}) for a
hard wall potential~($\gamma_{0}$), as well as numerical calculations
for step potentials with soft region widths of 15~nm~($\gamma_{1}$),
45~nm~($\gamma_{2}$) and 73~nm~($\gamma_{3}$). The top curve (h) shows
the calculation for the H\'enon-Heiles potential.  As in
Fig.~\protect\ref{simulations} the $\gamma_{2}$ potential matches the
data best together with a carrier density $n_{\rm A1} = 0.7\!\times\!
10^{15}$ m$^{-2}$.} 
\end{figure}

The presence of a single quasi-period over a range of magnetic fields and
gate voltages implies the existence of a single orbit which is robust
to both changes in magnetic field and the shape of the potential. By
analyzing the classical trajectories in soft and hard-walled model
potentials and homogeneous magnetic fields,
(see Sec.~\ref{sec:experiment}), we note that the triangular loop
orbit with the same winding direction as the free electron cyclotron
orbit is the only orbit which exists in the large field range of
the experimentally observed oscillations.  Rather than being destroyed
as the magnetic field is augmented, it gradually changes into the
cyclotron orbit which is also a single loop.  The triangular orbit at
low fields can therefore be seen as a cyclotron orbit forced to fit
inside the triangular cavity. Indeed, a detailed analysis of the
observed oscillations reveal a smooth transition from the low field
oscillations with a nearly constant quasi-period, to Landau level
quantization oscillations, or Shubnikov-de~Haas oscillations, periodic
in $B^{-1}$.  This is shown in Fig.~\ref{fielddependence}.  In panel
(a) is displayed a gray-scale plot of the conductance as a function of
both magnetic field $B$ and gate voltage $V_{\rm g}$.  It is clearly
seen how the resolved oscillations are of the $B^{-1}$ periodic
Shubnikov-de~Haas type in the open quantum dot ($V_{\rm g} \approx
-0.27$~V, top of panel (a)), where the  magnetic confinement
dominates the electrostatic confinement already when the first
oscillation can be resolved.  This is in contrast to the nearly closed
quantum dot ($V_{\rm g} \approx -0.55$~V, bottom of panel (a)), where
a more constant quasi-period is seen at low fields before the
$B^{-1}$ periodic behavior sets in. In the regime between the 
two limits ($G\sim e^2/h$ and $G \gg e^2/h$) where several conducting 
channels are passed through the system, the fluctuations are generally 
less clear, and at some gate voltages practically disappear.

In Figs.~\ref{fielddependence}b and~\ref{fielddependence}c the results 
of a quantitative analysis are shown.  We begin by considering the 
triangular periodic orbit in a hard-walled potential, where it in zero 
field is an equilateral triangle \cite{Brack97a}.  In weak fields, the
triangular orbit splits into two orbits: a large area orbit $(+)$
winding the same way as the cyclotron orbit, and a smaller area orbit
$(-)$ winding the opposite way.  Between the reflections at the walls
of the quantum dot, the electrons move on arcs with the cyclotron
radius $R_{\rm c} = \hbar k_{F}/eB$.  For weak magnetic fields a
$p$-corner periodic orbit exists in any regular $p$-sided polygonal
potential with hard walls, as for instance seen in
Ref.~\onlinecite{Bird96a}. In soft potentials, the corners of the
orbit will be rounded accordingly.  In terms of the dimensionless
magnetic field $b = l_{0}/2R_{c}$ the field dependent length $L(b)$
and area $F(b)$ of a $p$-corner periodic orbit, which at zero field
has sidelength $l_{0}$
\pagebreak
\noindent
can be written as

\begin{eqnarray}
        L(b) &=& \frac{pl_{0}}{b}\arcsin(b) \\
        F(b) &=& \frac{pl_{0}^2}{4}\left[
        \cot\left(\frac{\pi}{p}\right) +
        \frac{\arcsin(b)}{b^2} -
        \frac{\sqrt{1-b^2}}{b}
         \right].
        \label{arealength}
\end{eqnarray}

\noindent By inserting these equations into Eq.~(\ref{BS}) the
quantization rule becomes

\begin{equation}
        N_{\pm} =
        \frac{pl_{0}}{2\lambda_{\rm F}}
        \left[
\mp  b\cot\left(\frac{\pi}{\rm p}\right)
                + \frac{1}{b}\arcsin(b)
                 + \sqrt{1-b^2}
        \right],
        \label{BS2}
\end{equation}

\noindent where $N_{+}$ ($N_{-}$) is the orbit winding the same
(opposite) way as the cyclotron orbit.  The corresponding magnetic
field dependent quasi periods $\Delta B_{\pm}= |dN_{\pm}/dB|^{-1}$ are
then given by

\begin{eqnarray}
        \Delta B_\pm &=& \frac{4\Phi_{0}}{pl_{0}^2}
                \left|
                \mp \cot\left(\frac{\pi}{\rm p}\right)  -
                \frac{\arcsin(b)}{b^2}  +
                \frac{\sqrt{1-b^2}}{b}
                \right|^{-1}.
        \label{period}
\end{eqnarray}
By comparison with Eq.~(\ref{arealength}) it is seen that the quasi-period
of the $(+)$ orbit takes on the simple form $\Delta B_{+} =
\Phi_{0}/F(b)$.  Above a certain field strength $b^{*}=\sin(\pi/p)$,
where the cyclotron radius $R_{\rm c}$ is equal to the radius of the
circumscribed circle, the $(+)$ orbit is just the cyclotron orbit.  We
note that the Landau level filling factor $\nu = n\Phi_{0}/B$ can be
derived semiclassically by applying the Bohr-Sommerfeld quantization
rule Eq.~(\ref{BS}) to the cyclotron orbit, giving $N_{\rm LL}=\nu /2$
with the factor of $2$ due to spin degeneracy.  The corresponding
quasi-period $\Delta B_{\rm LL}$ is then quadratic in $b$ (see
Fig.~3):

\begin{equation}
\Delta B_{\rm LL} = \frac{2B^2}{n\Phi_{0}} = \frac{4\Phi_{0}}{\pi
l_{0}^2} b^2.
\label{sdhperiod}
\end{equation}
The $(-)$ orbit exists only at low fields $b<1$, and furthermore its
quasi-period diverges around 0.2 T, which disqualifies it as an
explanation for the observed oscillations.  We apply Eq.~(\ref{period})
for $b<b^{*}$ and Eq.~(\ref{sdhperiod}) for $b>b^{*}$ to our system by
setting $p=3$ (triangular orbit) and by using the geometrical relation
$l_{0} = (2-0.75\tilde{V})W$ for the zero-field orbit sidelength as a
function of the dimensionless gate voltage $\tilde{V}$ (see
Sec.~\ref{sec:experiment}).

In Fig.~\ref{fielddependence}b the calculated hard-wall
quasi-period denoted $\gamma_0$ is plotted against magnetic
field $B$ together with measured periods (black markers) of
conductance oscillations in sample A2 and similarly in
Fig.~\ref{fielddependence}c for sample A1. The measured
periods represent the distance between adjacent maxima (or minima) of
quasi-periodic oscillations plotted at the fields halfway between the
maxima (or minima).  The transition to a cyclotron orbit is taking
place at the sharp minimum of the curve.  In a more accurate
quantum-mechanical calculation, this minimum would be smoothed, as
the transition from one type of orbit to another is continuous due to
the finite extend of the wavefunction by which the hard wall
appears more soft \cite{Blaschke97a}. Likewise classically, the
hard-wall model predicts much larger variations of the quasi-period
$\Delta B(B)$ in the low field region, than obtained with softened
walls.  Therefore, to improve the hard-wall model we calculate the
quasi-period $|dN/dB|^{-1}$ numerically for the sequence of loop
orbits that exist in soft potentials in the field range from zero to
0.8 T. By this we not only obtain a closer match to the actual
potential, we also avoid the above mentioned discontinuities in the
transition from the triangular to the circular orbit.

In Fig.~\ref{fielddependence}b-c we plot in addition to the analytical
predictions of Eq.~(\ref{period}) and~(\ref{sdhperiod}), the periods
obtained numerically with the H\'enon-Heiles potential (h) and three
step potentials, $\gamma_1$ - $\gamma_3$, of different smoothing
widths $\gamma$.  Again we obtain the best agreement with the fairly
steep potentials softened only a little by choosing the same small
values of $\gamma$ and the same carrier density $n$ as in
Sec.~\ref{sec:classical}.

\begin{figure}[h]
\epsfysize=80mm \centerline{\epsfbox{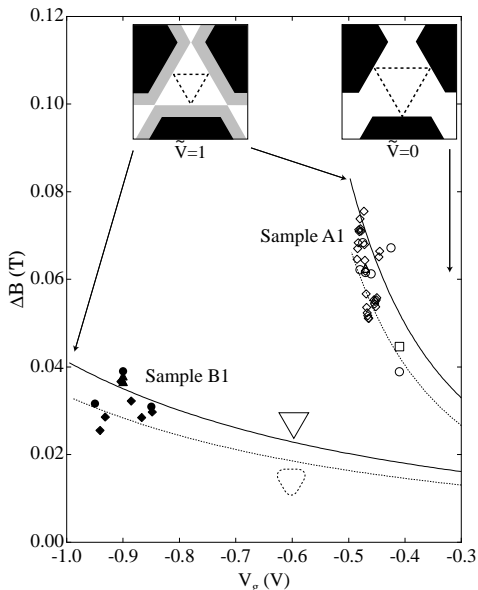}}
\caption{
\label{periods}
The measured quasi-periods determined from the power spectra of seven
measurement series plotted together with the calculated quasi-period
corresponding to the triangular orbits. For comparison, the areas of
the softened triangles, as estimated from numerical calculations, are
indicated by the dotted lines.}
\end{figure}

We note from the simulations that the soft walls tend to stabilize the
quasi-period of the $(+)$ orbit at low fields, and that the
simple relation $\Delta B_{+} = \Phi_{0}/F(b)$ found for hard walls
is approximately correct for softened walls as well. The field
dependence of the quasi-period related to the triangular orbit
then turns out to be similar to that of an orbit with constant area
and length.  These facts allow us to use the zero field triangular
orbit in the further analysis of oscillations at low fields.  With
the linear depletion model of the dot (see Sec.~\ref{sec:experiment})
we obtain the following approximative expression for the area of the
triangular orbit as a function of the dimensionless gate voltage
$\tilde{V}$:

\begin{equation}
   F_{\triangle}(\tilde{V})
   \simeq W^2(0.24\tilde{V}^2 - 1.30\tilde{V} + 1.73).
   \label{triangle}
\end{equation}
Eq.~(\ref{triangle}) contains no adjustable
parameters; it is based on the linear depletion model and the measured
data. In Fig.~\ref{periods} is shown 
that the theoretical prediction
of the oscillation periods $\Delta B = \Phi_0/F_{\triangle}$ as a
function of the gate voltage $V_{\rm g}$ is in good agreement with the
experimental data points. The fact that the measured periods are
slightly smaller than predicted is due to the hard wall triangle being
slightly smaller in area than its soft wall rounded counterpart. From
the simulations we estimated the difference in area to be roughly
20\%, which leads to the corrected curves (dashed) in
Fig.~\ref{periods}.

Finally, we observe an almost complete suppression of random
fluctuations in the oscillatory regime, which is clear from the
isolated peaks in the power spectra of Fig.~\ref{oscillations}.  A
possible factor to the suppression of random fluctuations is the small
ratio between size and wavelength caused by the
low carrier densities.  When sample A1 is smallest (near pinch-off),
the length of the symmetry axis from gate to lead is about 6 times the
Fermi wavelength $\lambda_{\rm F}=80$ nm, while the total length $L$
of the triangular orbit is 10~$\lambda_{\rm F}$, compared to roughly
18~$\lambda_{\rm F}$ for the square orbit considered in
Ref.~\onlinecite{Bird96a}.  The small sample A1 does not quite fulfill
the condition $\lambda_{\rm F}\ll L$ and random fluctuations are
suppressed.

\section{Discussion}
\label{sec:discussion}

Having established that the low field conductance oscillations are
due to the triangular periodic orbit, which changes into a cyclotron
orbit at high fields,  we discuss qualitatively the coupling between
the incoming electrons and the orbit causing the conductance
oscillations.

\begin{figure}[b]
\epsfxsize=\columnwidth \centerline{\epsfbox{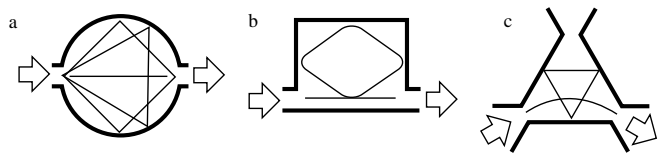}}
\caption{
\label{orbits}
Panels (a) and (b) show the geometries discussed by
Persson\protect\cite{Persson95a} and Bird\protect\cite{Bird96a} (see
text).  Some periodic orbits related to the conductance oscillations
observed in the samples are sketched.  In both cases the collimated
beam can inject close to a periodic orbit in phase space.  In panel c
the triangular sample is sketched.  The incoming orbit does not match
the direction of the triangular orbit anywhere; this orbit is
classically inaccessible in phase space, in contrast to the
situation shown in panel (a) and (b)}
\end{figure}

Recently, quasiperiodic magnetoconductance oscillations in square and
triangular quantum dots have been interpreted in terms of selective
probing\cite{Bird96a,Zozoulenko97a,Christensson97a}  where the incoming
electrons are injected close to the classically periodic orbits as
sketched in Fig.~\ref{orbits}a and~\ref{orbits}b. The same
interpretation may in fact also be applied to circular.
dots\cite{Marcus92a,Persson95a,Berry94a} A common feature of these
billiard shapes is that the polygons accessible from the leads are
single members of families of degenerate orbits with the same
classical action. As an example, in the circular case these are
rotated copies of the accessed polygon. Classically, small-angle
scattering can transfer a particle from one orbit (or family) to
another until it finally couples to the lead and escape is
possible. The situation for our triangular samples is entirely
different: the orientation of the triangular loop orbit is fixed with
respect to the leads, and the orbit is isolated, i.e.\ it does not
belong to any continuous family of orbits.  As illustrated in
Fig.~\ref{orbits}c, the incoming electron beam does not inject into the
loop orbit.  From the analysis of the classical dynamics, we have
noted that just a small fraction of injected electrons end up in
trajectories close to the loop orbit at small fields, 
$B<0.15$~T, and
none at higher fields, where the loop orbit is localized in the center
of the dot.  So even if the quasi-period of the oscillations can be
explained in terms of one periodic orbit, the coupling between the
incoming trajectories and the periodic orbit is expected a priori to
be weak. But we will argue that the remarkable robustness and strength
of the observed oscillations is exactly a consequence of this weak
coupling. In a sense the triangular orbit resembles a resonating
oscillator only weakly coupled to the surroundings, but the weak
coupling translates into a high $Q$-value with a significant response
as a consequence in steady-state.

Quantum-mechanically the spatial structure of high-lying eigenstates
often resembles periodic orbits of the corresponding classical system.
In the semiclassical limit, i.e.\ for mesoscopic systems,
these so-called scarred wavefunctions are particularly
important\cite{Bird96a,Zozoulenko97a,Christensson97a,SemiclReview}.
We speculate that the triangular orbit in our system is largely
represented by a set of scarred wavefunctions similar to those found
also in a triangular quantum dot \cite{Christensson97a}. In that
study, however, a different lead geometry implies a strong coupling
between the periodic orbits and the leads. It is possible to explain
the weak coupling found in our classical by noting that
the propagation direction of the incoming wave will be almost
orthogonal with the scarred wavefunction propagating along the
triangular orbit thus yielding an almost vanishing overlap integral.
The experimental evidence of a non-zero coupling would then be a
manifestation of quantum mechanical tunneling in phase space without
classical counterparts \cite{Persson95a,Bohigas}. A similar situation
has been seen in the high field Aharonov-Bohm effect in quantum
dots\cite{Wees89a}, where the outer edge state propagates through the
dot while the inner edge state is an isolated loop.  In this case,
tunneling between the outer and inner state leads to
magnetoconductance oscillations corresponding to the area of the inner
isolated state. In the wave picture, as in the classical picture, the
remarkable robustness and strength of the observed oscillations can be
explained in terms of high $Q$-values: once a wave packet is launched
along the isolated triangular orbit, small-angle scattering cannot
bring it close to an exit lead; only a tunneling process in phase
space can do so. The triangular orbit acts as a wave resonator with a
high $Q$-value, and we come to the surprising conclusion that periodic
orbits can be observed most clearly in those transport measurements where
the orbits in question are not directly accessible the incoming
particles injected from the leads but in fact are only weakly coupled
to the leads.

As a final remark we note that the smooth transition from
oscillations in $B$ to oscillations in $1/B$ indicates that the low
field oscillations is a density of states effect similar to Landau
quantization, and that the main difference of the two field regimes is
whether the confinement is of electrostatic or magnetic origin. Loosely
speaking, the density of states from the inner dot contributes
significantly to the scattering resonances, and the conductance
oscillations can be interpreted as a manifestation of gross-shell
structure in the energy spectrum related to the existence of periodic
orbits\cite{SemiclReview}.
A consequence of the appearance of such a gross-shell structure would
be that the thermal behavior of the fluctuation amplitude differs
from those of chaotic energy spectra.  Due to level repulsion in a
chaotic system, the typical level spacing is equal to the average
level spacing $\langle \Delta \rangle$.  The amplitude of the
fluctuations increases with decreasing temperature, only until the
thermal energy $k_{\rm B}T$ becomes smaller than $\langle \Delta
\rangle$, where the number of available scattering states becomes
independent on the temperature. This has been seen experimentally in a
system similar to ours\cite{Bird95c}.  However, for systems showing
gross-shell structure the typical level spacing  becomes smaller than
$\langle \Delta \rangle$ due to the appearance of large gaps
$\Delta_{\rm gap}>\langle \Delta \rangle$ in the energy spectrum.
Saturation should then set in on the larger energy scale $\Delta_{\rm
gap}$, giving rise to a higher saturation temperature $T \approx
\Delta_{\rm gap}/k_{\rm B}$.

In the smallest triangular sample A1, we observe indeed such a
saturation of $\langle \delta G \rangle$ below temperatures of 0.5~K
and corresponding bias voltages $eV_{\rm bias}/k_{\rm B}$. This effect
is not due to trivial saturation of the electron gas temperature,
which only occurs at much lower temperature ($\sim 0.05$~K). This is
evident not only from resistance measurements in the fractional
quantum Hall regime (see Sec.~\ref{sec:experiment}) but also from the
temperature dependence of the conductance fluctuations: the shape of
the conductance fluctuations changes throughout the temperature range
from 0.5~K to 0.05~K, and simultaneously the correlation field 
as obtained from the autocorrelation of the fluctuations decreases 
below the transition temperature. However, applying the same analysis
as in Ref.~\onlinecite{Bird95c} to our data we obtain saturation
energies 1.5 to 4 times larger than the expected value $\langle \Delta
\rangle$ indicating the presence of the above mentioned shell
structure. Furthermore, we observed a tendency of the saturation to
set in at higher temperatures, when the dot was near pinch off,
consistent with an increased shell spacing  as the dot is made smaller.
If this explanation is correct, the temperature dependence of the
fluctuations provides an interesting possibility of detecting the
quantum properties of small cavities with mixed dynamics.  We propose
that the bias voltage could be used as a convenient parameter to
control the thermal broadening of the levels in order to extract
information on the level distribution as described above.

\section{Summary}
\label{sec:summary}

We have studied the intermediate and low temperature
magnetoconductance of triangular symmetric quantum dots of different
sizes.  At low temperatures we observe strong periodic fluctuations in
the magnetoconductance, which are robust to moderate variations in
gate voltage and magnetic field.  While the quasi-period is
nearly constant at low magnetic fields, a $1/B$ periodicity is
observed at higher fields, similar to Shubnikov-de~Haas oscillations.
The magnetic field dependence of the oscillation periods has been
analyzed numerically and analytically in terms of one classical orbit:
the triangular loop orbit. To account for the effect of gate
voltage on the area of the orbit we assumed linear depletion as a
function of gate voltage.  This leads to a geometrical model without
adjustable parameters that agrees well with the measured oscillation
periods obtained at different gate voltages.  The orbit in question
is inaccessible in terms of classical collimated trajectories, and
we therefore propose that the transport mechanism is a pure quantum
tunneling effect, through the essentially isolated triangular
periodic orbit. The quasi-isolation of the orbit explains the
robustness and strength of the observed oscillations, and our work has
lead us to the conclusion that periodic orbits are observed very
clearly in systems where the orbits are not directly accessible the
incoming particles injected from the leads. Finally, the
quasi-isolation of the triangular orbit leads to the appearance of
shell effects with implications for the saturation temperature of
the quantum fluctuations: we have measured a saturation temperature up
to four times larger than the one expected for systems without shell
structure.

\section*{Acknowledgements}

We are particularly indebted to M.~Persson, who performed the
excellent e-beam exposure, and provided the initial motivation to
begin this study, and to C. B. S\o rensen for producing high quality
sample material at NANOLAB. We thank M.~Brack, K.~Astrup Eriksen,
P.~Hedeg\aa rd and C.M.~Marcus for helpful discussions and  R.~Jensen
for technical assistance.

This work was supported by CNAST. H.~Bruus was supported by the Danish
Natural Science Research Council through Ole R\o mer grant
no. 9600548, and S.M. Reimann acknowledges financial support by the
Studienstiftung des deutschen Volkes and the BASF AG.


\end{document}